\documentclass[11pt]{article}
\usepackage{latexsym}
\usepackage{amssymb}
\usepackage[dvips]{graphicx}
\usepackage{epsfig}
\usepackage{rotating}
\usepackage{amsfonts}
\usepackage{amsmath}
\begin{document}

\begin{titlepage}
\begin{flushright}
CP3-09-38\\
\end{flushright}

\vspace{20pt}

\begin{center}

{\Large\bf The ${\cal N}=1$ Supersymmetric Landau Problem}\\

\vspace{5pt}

{\Large\bf and its Supersymmetric Landau Level Projections:}\\

\vspace{5pt}

{\Large\bf the ${\cal N}=1$ Supersymmetric Moyal--Voros Superplane}

\vspace{20pt}

Joseph Ben Geloun$^{a,c,d,*}$, Jan Govaerts$^{b,c,\dag,}\footnote{Fellow of the Stellenbosch Institute
for Advanced Study (STIAS), 7600 Stellenbosch, South Africa.}$ and Frederik G. Scholtz$^{a,\ddag}$

\vspace{15pt}

$^{a}${\sl National Institute for Theoretical Physics (NITheP),\\
Private Bag X1, Matieland 7602, South Africa}\\
\vspace{10pt}
$^{b}${\sl Center for Particle Physics and Phenomenology (CP3),\\
Institut de Physique Nucl\'eaire, Universit\'e catholique de Louvain (U.C.L.),\\
2, Chemin du Cyclotron, B-1348 Louvain-la-Neuve, Belgium}\\
\vspace{10pt}
$^{c}${\sl International Chair in Mathematical Physics and Applications (ICMPA--UNESCO Chair),\\
University of Abomey--Calavi, 072 B. P. 50, Cotonou, Republic of Benin}\\
\vspace{10pt}
$^{d}${\sl D\'epartement de Math\'ematiques et Informatique,\\
Facult\'e des Sciences et Techniques, Universit\'e Cheikh Anta Diop, Senegal}

\vspace{20pt}

E-mail:  $^{*}${\em bengeloun@sun.ac.za},\quad $^{\dag}${\em Jan.Govaerts@uclouvain.be}, 
\quad $^{\ddag}${\em fgs@sun.ac.za } 


\vspace{10pt}

\begin{abstract}
\noindent
The ${\cal N}=1$ supersymmetric invariant Landau problem is constructed and solved.
By considering Landau level projections remaining non trivial under ${\cal N}=1$
supersymmetry transformations, the algebraic structures of the ${\cal N}=1$ supersymmetric
covariant non(anti)commutative superplane analogue of the ordinary ${\cal N}=0$
noncommutative Moyal--Voros plane are identified.
\end{abstract}

\end{center}

Pacs number 11.10.Nx

\end{titlepage}

\setcounter{footnote}{0}

\section{Introduction}
\label{Intro}

Over the years, the classic textbook example\cite{jackiw} of the quantum Landau problem has remained a constant source of fascination
and inspiration, in fields apparently so diverse as condensed matter physics, the fundamental unification of gravity with high
energy quantum physics, or purely mathematical studies in noncommutative deformations of geometry. The quantum Landau system with
its Landau level structure of energy quantum states provides a natural model for the integer and fractional quantum
Hall effects, whether in its commutative or noncommutative formulations\cite{suss,jel}. The noncommutative geometry resulting from
a projection onto any of the Landau levels\cite{macris,scholtz} provides the basic example of deformation quantisation through the
Moyal--Weyl (or Voros) $*$-product\cite{scholtz2}. The same algebraic structures are also realised in M-theory
for specific background field configurations\cite{SW}.

By accounting for the spin degrees of freedom of the charged particle in the Landau problem---after all this is the physical
situation for spin $1/2$ electrons---raises the possibility of
extending further such algebraic structures in a manner consistent with supersymmetry. Through appropriate
projections onto Landau levels, one wonders then how the noncommutative Euclidean Moyal--Voros plane (or torus) extends
into a Grassmann graded noncommutative variety, whose originally commuting and anticommuting variables now
possess some deformed algebra representative of a non(anti)commutative Grassmann graded geometry with supersymmetry.

Extensions of the Landau problem including Grassmann graded degrees of freedom and algebraic structures of the supersymmetric type
have been considered over the years, whether in a planar, toroidal or spherical geometry\cite{ivanov}, but apparently never explicitly with
the above purpose in mind. Deformations of the algebraic structures of even a Grassmann graded Landau problem have so far not
led to a projection onto Landau levels on which supersymmetry acts non trivially\cite{BGH,bgh}. We note also that the supersymmetry
inherent to the Landau problem with a spin $1/2$ charged particle of gyromagnetic factor $g=2$ does not survive a noncommutative
geometry on the Euclidean plane\cite{BenGeloun}.

With the above condensed matter and mathematical physics contexts in the back of one's mind as motivations for the potential
relevance of such an analysis, in the present work we present a first investigation addressing the question raised above, restricted
to the simplest case of a single supersymmetry, ${\cal N}=1$. Results for a larger number of supersymmetries are to be discussed
elsewhere\cite{BenGeloun2}. Naively one might expect that besides the non-commutative coordinates of the Euclidean Moyal--Voros plane,
the non-anticommutative Grassmann odd sector may amount simply to some rescaling of the associated Clifford algebra, and retain its
commutativity with the Grassmann even sector. Surprisingly perhaps, this is not what our analysis reveals. Rather,
in what may as well be called the ${\cal N}=1$ supersymmetric Moyal--Voros superplane, there
is a transmutation of sorts, between the fermionic and some of the bosonic degrees of freedom, resulting in degrees of
freedom of fermionic character and yet possessing an integer spin. The spin $1/2$ degrees of freedom end up coinciding with
one of the chiral modes of the bosonic sector, and vice-versa, such that the latter obeys a fermionic statistics nonetheless.
Finally these bosonic-fermionic coordinates solder two copies of the ordinary ${\cal N}=0$ Moyal--Voros plane, to
produce its ${\cal N}=1$ superplane extension.

In Section 2, the results of the ordinary Landau problem relevant to our analysis are recalled.
Section 3 then considers the general problem of a ${\cal N}=1$ supersymmetric extension of the nonrelativistic
particle coupled to an arbitrary background magnetic field, in an Euclidean space of whatever dimension $d$.
When particularised to the plane with a static and homogeneous magnetic field perpendicular to that plane,
the ${\cal N}=1$ supersymmetric Landau problem is solved explicitly. Section 4 then considers projections
onto Landau levels preserving the ${\cal N}=1$ supersymmetry, and identifies the algebraic structures
resulting from such projections which characterise the ${\cal N}=1$ supersymmetric Moyal--Voros superplane,
with in particular the boson-fermion transmutation mentioned above.
Section 5 presents the final and complete form of the ${\cal N}=1$ supersymmetric Moyal--Voros superplane,
as well as some concluding remarks.

\section{The Ordinary Landau Problem}
\label{Sect2}

If only to set our notations and emphasize some specific points to be contrasted with the situation in the
${\cal N}=1$ case, even though this is standard material let us first reconsider the well known ordinary problem
of a charged particle of mass $m$ moving in the Euclidean plane of Cartesian coordinates $(x_1,x_2)$ and subjected
to a static and homogeneous magnetic field perpendicular to that plane. In the symmetric gauge for the corresponding
vector potential, the action principle of that system is defined in terms of the following Lagrangian,
\begin{equation}
L_0=\frac{1}{2}m\left(\dot{x}^2_1+\dot{x}^2_2\right)-\frac{1}{2}B\left(\dot{x}_1 x_2-\dot{x}_2 x_1\right).
\end{equation}
In this expression, $B$ stands for the magnetic field with a normalisation which includes the charge of
the particle. Furthermore, without loss of generality it is assumed that the (right-hand) orientation of the plane $(x_1,x_2)$
is such that $B>0$ (as usual a dot above a quantity stands for its first-order time derivative).

The Hamiltonian formulation of the dynamics is obtained as follows. Besides the momenta $(p_1,p_2)$ canonically
conjugate to the configuration space variables $(x_1,x_2)$, respectively, such that
\begin{eqnarray}
p_1 &=& \frac{\partial L_0}{\partial\dot{x}_1}=m\dot{x}_1-\frac{1}{2}Bx_2,\qquad
\dot{x}_1=\frac{1}{m}\left(p_1+\frac{1}{2}Bx_2\right), \nonumber \\
p_2 &=& \frac{\partial L_0}{\partial\dot{x}_2}=m\dot{x}_2+\frac{1}{2}Bx_1,\qquad
\dot{x}_2=\frac{1}{m}\left(p_2-\frac{1}{2}Bx_1\right),
\end{eqnarray}
time evolution is generated through canonical Poisson brackets, $\left\{x_i,p_j\right\}=\delta_{ij}$, from the canonical Hamiltonian,
\begin{eqnarray}
H_0 &=& \dot{x}_1 p_1 + \dot{x}_2 p_2 - L_0 =
 \frac{1}{2m}\left(p_1+\frac{1}{2}Bx_2\right)^2+\frac{1}{2m}\left(p_2-\frac{1}{2}Bx_1\right)^2 \nonumber \\
&=& \frac{1}{2m}\left(p^2_1+p^2_2\right)+\frac{1}{2}m\left(\frac{\omega_c}{2}\right)^2\left(x^2_1+x^2_2\right)
-\frac{1}{2}\omega_c\left(x_1 p_2 - x_2 p_1\right),
\label{eq:H0}
\end{eqnarray}
where the following notation for the cyclotron angular frequency is introduced,
\begin{equation}
\omega_c=\frac{B}{m}.
\end{equation}
One reason for the above choice of the circular gauge is that it makes manifest the invariance of the dynamics under SO(2)
rotations in the plane. The Noether generator of these transformations in infinitesimal form reads,
\begin{equation}
L\equiv L_{\rm Noether}=x_1 p_2 - x_2 p_1,
\label{eq:L}
\end{equation}
which besides the total energy $H_0$, is a second and independent constant of the motion, namely the total angular-momentum of the
system inclusive of the magnetic field contribution. This latter fact implies that the ordinary Landau problem (which does
not include further interactions with some potential energy, $V(x_1,x_2)$) is a two degrees of freedom system
which is integrable in the Liouville sense.

\subsection{Landau levels}

The quantum Landau problem is thus defined by the Heisenberg algebra of hermitian operators $(x_1,p_1)$ and $(x_2,p_2)$,
\begin{equation}
\Big[x_i,p_j\Big]=i\hbar\delta_{ij}\mathbb{I},\qquad x^\dagger_i=x_i,\qquad p^\dagger_i=p_i,\qquad i,j=1,2,
\end{equation}
as well as the quantum Hamiltonian and angular-momentum given by the expressions in (\ref{eq:H0}) and (\ref{eq:L})
(for which no operator ordering ambiguities arise. Note well that quantum operators are not distinguished from their classical couterparts
by using a ``hat" notation. There is no risk of confusion given the context).

Introducing first the Cartesian Fock algebra operators,
\begin{equation}
a_i=\frac{1}{2}\sqrt{\frac{B}{\hbar}}\left(x_i+\frac{2i}{B}p_i\right),\qquad
a^\dagger_i=\frac{1}{2}\sqrt{\frac{B}{\hbar}}\left(x_i-\frac{2i}{B}p_i\right),
\end{equation}
with,
\begin{equation}
\left[a_i,a^\dagger_j\right]=\delta_{ij}\,\mathbb{I},
\end{equation}
one has,
\begin{equation}
x_i=\sqrt{\frac{\hbar}{B}}\left(a_i + a^\dagger_i\right),\qquad
p_i=-\frac{1}{2}iB\sqrt{\frac{\hbar}{B}}\left(a_i - a^\dagger_i\right).
\end{equation}

Next consider the chiral or helicity Fock algebra operators,
\begin{equation}
a_\pm=\frac{1}{\sqrt{2}}\left(a_1 \mp i a_2\right),\qquad
a^\dagger_\pm=\frac{1}{\sqrt{2}}\left(a^\dagger_1 \pm i a^\dagger_2\right),
\end{equation}
with
\begin{equation}
\left[a_\pm, a^\dagger_\pm\right]=\mathbb{I},\qquad
\left[a_\pm, a^\dagger_\mp\right]=0,
\end{equation}
and such that,
\begin{eqnarray}
a_1 &=& \frac{1}{\sqrt{2}}\left(a_+ + a_-\right),\qquad
a^\dagger_1=\ \ \frac{1}{\sqrt{2}}\left(a^\dagger_+ + a^\dagger_-\right), \nonumber \\
a_2 &=& \frac{i}{\sqrt{2}}\left(a_+ - a_-\right),\qquad
a^\dagger_2 = -\frac{i}{\sqrt{2}}\left(a^\dagger_+ - a^\dagger_-\right).
\end{eqnarray}
In terms of these the phase space operators read
\begin{eqnarray}
x_1 &=& \ \sqrt{\frac{\hbar}{2B}}\left( a_+ + a_- + a^\dagger_+ + a^\dagger_- \right),\qquad
p_1 = -i\frac{B}{2}\sqrt{\frac{\hbar}{2B}}\left(a_+ + a_- - a^\dagger_+ - a^\dagger_- \right), \nonumber \\
x_2 &=& i\sqrt{\frac{\hbar}{2B}}\left(a_+ - a_- - a^\dagger_+ + a^\dagger_-\right),\qquad
p_2=\frac{B}{2}\sqrt{\frac{\hbar}{2B}}\left(a_+ - a_- + a^\dagger_+ - a^\dagger_-\right).
\label{eq:invert}
\end{eqnarray}
In particular,
\begin{equation}
p_1+\frac{1}{2}Bx_2=-iB\sqrt{\frac{\hbar}{2B}}\left(a_- - a^\dagger_-\right),\qquad
p_2-\frac{1}{2}Bx_1=-B\sqrt{\frac{\hbar}{2B}}\left(a_- + a^\dagger_-\right),
\label{eq:pair1}
\end{equation}
\begin{equation}
p_1-\frac{1}{2}Bx_2=-iB\sqrt{\frac{\hbar}{2B}}\left(a_+ - a^\dagger_+\right),\qquad
p_2+\frac{1}{2}Bx_1=B\sqrt{\frac{\hbar}{2B}}\left(a_+ + a^\dagger_+\right),
\label{eq:pair2}
\end{equation}
thus showing that up to some normalisation factor, each of these two pairs of operators
plays a r\^ole analogous to a Heisenberg algebra for each of the chiral sectors separately.

In terms of the chiral Fock algebra operators, one finds,
\begin{equation}
H_0 = \hbar\omega_c a^\dagger_- a_- + \frac{1}{2}\hbar\omega_c,\qquad
L = \hbar\left(a^\dagger_+ a_+ - a^\dagger_- a_-\right),
\end{equation}
with in particular,
\begin{equation}
\left[L,a_\pm\right]=\mp\hbar\, a_\pm,\qquad
\left[L,a^\dagger_\pm\right]=\pm\hbar\, a^\dagger_\pm,
\end{equation}
thus showing that the creation operator $a^\dagger_+$ (resp., $a^\dagger_-$) creates a quantum
carrying a unit $(+\hbar)$ (resp., $(-\hbar)$) of angular-momentum.

The orthonormalised chiral Fock state basis with as normalised Fock vacuum a state $|\Omega\rangle\equiv|0,0\rangle$ such that
\begin{equation}
a_\pm|\Omega\rangle=0,\qquad \langle\Omega|\Omega\rangle=1,
\end{equation}
is constructed by
\begin{equation}
|n_+,n_-\rangle=\frac{1}{\sqrt{n_+!\,n_-!}}\left(a^\dagger_+\right)^{n_+}\left(a^\dagger_-\right)^{n_-}|\Omega\rangle,\qquad
\langle n_+,n_-|m_+,m_-\rangle=\delta_{n_+,m_+}\,\delta_{n_-,m_-},
\end{equation}
with the property that
\begin{equation}
\sum_{n_+,n_-=0}^\infty|n_+,n_-\rangle\,\langle n_+,n_-|=\mathbb{I}.
\end{equation}
This complete set of states is a basis which diagonalises both commuting operators $H_0$ and $L$,
\begin{equation}
H_0|n_+,n_-\rangle=\hbar\omega_c\left(n_- + \frac{1}{2}\right)\,|n_+,n_-\rangle,\qquad
L|n_+,n_-\rangle=\hbar\left(n_+ - n_-\right)\,|n_+,n_-\rangle .
\end{equation}
For any given value of $N=0,1,2,\ldots$, the set of states $|n_+,N\rangle$ ($n_+=0,1,2,\ldots$)
is thus infinite countable degenerate, spanning the Landau sector at level $N$ with energy $\hbar\omega_c(N+1/2)$.
This degeneracy is lifted by adding to the system some interaction potential energy, $V(x_1,x_2)$.

Note that given the above resolution of the unit operator, the chiral annihilation and creation operators have
the following representations,
\begin{equation}
a_-=\sum_{n_+,n_-=0}^\infty |n_+,n_-\rangle\ \sqrt{n_-+1}\ \langle n_+,n_-+1|,\qquad
a^\dagger_-=\sum_{n_+,n_-=0}^\infty |n_+,n_-+1\rangle\ \sqrt{n_-+1}\ \langle n_+,n_-|,
\end{equation}
\begin{equation}
a_+=\sum_{n_+,n_-=0}^\infty |n_+,n_-\rangle\ \sqrt{n_++1}\ \langle n_++1,n_-|,\qquad
a^\dagger_+=\sum_{n_+,n_-=0}^\infty |n_++1,n_-\rangle\ \sqrt{n_++1}\ \langle n_+,n_-|.
\end{equation}

\subsection{Landau level projections}

Let us now consider a fixed Landau sector at level $N=0,1,2,\ldots$ and the associated
projection operator,
\begin{equation}
\mathbb{P}_N=\sum_{n_+=0}^\infty |n_+,N\rangle \langle n_+,N|,\qquad
\mathbb{P}^\dagger_N=\mathbb{P}_N,\qquad
\mathbb{P}^2_N=\mathbb{P}_N.
\end{equation}
Given any operator $A$, let us denote by $\overline{A}$ the associated operator projected onto
the Landau sector at level $N$, $\overline{A}=\mathbb{P}_N\,A\,\mathbb{P}_N$.

In the particular case of the chiral Fock operators, one finds,
\begin{equation}
\overline{a}_-=0,\qquad \overline{a^\dagger_-}=\overline{a}^\dagger_-=0,
\end{equation}
\begin{equation}
\overline{a}_+=\sum_{n_+=0}^\infty |n_+,N\rangle\ \sqrt{n_++1}\, \langle n_++1,N|,\qquad
\overline{a^\dagger_+}=\overline{a}^\dagger_+=\sum_{n_+=0}^\infty |n_++1,N\rangle\ \sqrt{n_++1}\ \langle n_+,N|,
\end{equation}
and thus,
\begin{equation}
\overline{a}_+\overline{a}^\dagger_+=\sum_{n_+=0}^\infty |n_+,N\rangle\ \left(n_++1\right)\ \langle n_+,N|,\qquad
\overline{a}^\dagger_+\overline{a}_+=\sum_{n_+=0}^\infty |n_+,N\rangle\ n_+\ \langle n_+,N|.
\end{equation}
Consequenly the projected right-handed chiral Fock operators still span a Fock algebra on the Landau level $N$,
\begin{equation}
\Big[\overline{a}_+,\overline{a}^\dagger_+\Big]=\mathbb{P}_N,
\end{equation}
while the left-handed chiral modes $a^{(\dagger)}_-$ are simply projected onto the null operator.

When considering the original phase space operators $(x_i,p_i)$, these results imply that
\begin{eqnarray}
\overline{x}_1 &=&\ \sqrt{\frac{\hbar}{2B}}\left(\overline{a}_+ + \overline{a}^\dagger_+\right),\qquad
\overline{p}_1 =-i\frac{B}{2}\sqrt{\frac{\hbar}{2B}}\left(\overline{a}_+ - \overline{a}^\dagger_+\right)
=-\frac{1}{2}B\overline{x}_2, \nonumber \\
\overline{x}_2 &=&i\sqrt{\frac{\hbar}{2B}}\left(\overline{a}_+ - \overline{a}^\dagger_+\right),\qquad
\overline{p}_2=\ \ \ \frac{B}{2}\sqrt{\frac{\hbar}{2B}}\left(\overline{a}_+ + \overline{a}^\dagger_+\right)
=\ \ \,\frac{1}{2}B\overline{x}_1.
\end{eqnarray}
Among the two pairs of conjugate operators in (\ref{eq:pair1}) and (\ref{eq:pair2}), upon projection
one vanishes identically\footnote{Note that since the pair which vanishes is precisely the one that contributes
in the Hamiltonian $H_0$, this fortuitous circumstance is such that projection onto the Landau level $N$ is equivalent
to taking a massless limit of the initial system, $m\rightarrow 0^+$, provided the quantum energy $\hbar\omega_c(N+1/2)$
of that Landau level is first subtracted from the Hamiltonian in order that all Landau levels decouple but for the one
under consideration.}. Or equivalently, projection onto any given Landau level implies that the
conjugate momentum operators $p_i$ are no longer independent from the Cartesian coordinate operators $x_i$,
and become proportional to these. As a consequence, the latter no longer commute as they do when
acting on the set of all Landau levels. Rather, on the Landau level $N$ one now finds,
\begin{equation}
\Big[\overline{x}_1,\overline{x}_2\Big]=-i\frac{\hbar}{B}\mathbb{P}_N.
\label{eq:NC0}
\end{equation}
Through projection onto the Landau level $N$, the four dimensional phase space of the system has been
projected to a two dimensional phase space with as conjugate pair of variables the two (projected) Cartesian coordinates
of the plane obeying, up to some normalisation factor, the usual Heisenberg algebra. The original commuting
algebra spanned by the operators $x_1$ and $x_2$ is deformed into a noncommutative algebra spanned by the
operators $\overline{x}_1$ and $\overline{x}_2$. These latter two operators define the noncommutative Moyal--Voros
plane characterised by the non vanishing commutator (\ref{eq:NC0}). From the present point of view, the noncommutative
Moyal--Voros plane is seen to be nothing else than the representation space of a single Fock algebra, in the present
instance that of the right-handed chiral Fock algebra restricted to the Landau level $N$. In effect, this algebra
also corresponds to the noncommutative quantum phase space of a one degree of freedom system obeying the Heisenberg algebra.

The projected Hamiltonian operator reduces to,
\begin{equation}
\overline{H}_0=\hbar\omega_c\left(N+\frac{1}{2}\right)\,\mathbb{P}_N,
\end{equation}
namely simply a multiple of the unit operator acting on the Landau level $N$. For the angular-momentum
operator one has
\begin{equation}
\overline{L}=\sum_{n_+=0}^\infty|n_+,N\rangle\,\hbar n_+\ \langle n_+,N|\,-\,\hbar N\mathbb{P}_N=
\hbar\left(\overline{a}^\dagger_+\overline{a}_+ - N\mathbb{P}_N\right).
\end{equation}
In particular,
\begin{equation}
\Big[\overline{L},\overline{a}_+\Big]=-\hbar\,\overline{a}_+,\qquad
\left[\overline{L},\overline{a}^\dagger_+\right]=+\hbar\,\overline{a}^\dagger_+,
\end{equation}
showing that the projected right-handed chiral creation operator $\overline{a}^\dagger_+$ still creates quanta carrying
a unit $(+\hbar)$ of angular-momentum. Quanta of angular-momentum unit $(-\hbar)$ however, can no longer be created.
The background magnetic field breaks time reversal invariance and selects a chiral orientation of the plane,
such that states of only one chirality are retained when restricted to a given Landau sector.

As a matter of fact, one could envisage the possibility of projecting the quantum Landau problem onto
a collection of $(M+1)$ successive Landau levels $(M=0,1,2,\ldots$) , in terms of a projector,
\begin{equation}
\mathbb{P}_{N,M}=\mathbb{P}_N+\mathbb{P}_{N+1}+\cdots+\mathbb{P}_{N+M},
\end{equation}
the value $M=0$ corresponding to the discussion above. However, such a procedure does not appear to offer
any particular advantage (in the absence of supersymmetry at this stage), and we shall not consider it any further here.
Let us only point out that under such a generalised projection the left-handed chiral operators $a^{(\dagger)}_-$
are no longer projected to the null operator. They rather have the following algebra,
\begin{equation}
\Big[\overline{a}_-,\overline{a}^\dagger_-\Big]=
\left(N+1\right)\mathbb{P}_N+\mathbb{P}_{N+1}+\cdots+\mathbb{P}_{N+M-1}-\left(N+M\right)\mathbb{P}_{N+M},
\end{equation}
while for the projected right-handed operators we still have a Fock algebra,
\begin{equation}
\Big[\overline{a}_+,\overline{a}^\dagger_+\Big]=\mathbb{P}_{N,M}.
\end{equation}
In particular, this implies for the projected Cartesian coordinates of the plane,
\begin{equation}
\Big[\overline{x}_1,\overline{x}_2\Big]=-\frac{i\hbar}{B}
\Big[-N\mathbb{P}_N+\left(N+M+1\right)\mathbb{P}_{N+M}\Big].
\end{equation}
Note well however that when $M\ne 0$ phase space is not reduced, and one still has to consider
the complete set of noncommuting operators $\overline{x}_i$ and $\overline{p}_i$ ($i=1,2$)  as independent ones.

Incidentally, these results are at odds with those in Ref.\cite{magro} as far as projection
onto more than a single Landau level is concerned.

Finally let us point out that the above construction involving projection onto any given Landau sector
of the full Hilbert space as such, is in no way related to the choice of Hamiltonian. Rather, given the
parameter $B$ in combination with $\hbar$, out of the $(x_i,p_i)$ Heisenberg algebra it is always possible to define
the chiral Fock algebras for $(a_\pm,a^\dagger_\pm)$, hence the Landau sectors $|n_+,N\rangle$ for a fixed $N=0,1,2,\ldots$
It only so happens that the chiral Fock states $|n_+,n_-\rangle$ also diagonalise the operators $H_0$
and $L$. Had the Hamiltonian operator been different from $H_0$ (for instance by adding an interaction
potential energy to it, $V(x_1,x_2)$), its eigenspectrum would not have coincided with the chiral Fock
state basis $|n_+,n_-\rangle$. Still, independently from this one may consider the different projections
onto Landau sectors considered in this Section.

\section{The ${\cal N}=1$ Supersymmetric Landau Problem}

Before considering the ${\cal N}=1$ supersymmetric Landau problem {\it per se\/}, let us construct
the action for the ${\cal N}=1$ supersymmetric nonrelativistic particle coupled
to an arbitrary magnetic field in an Euclidean space of arbitrary dimension $d\ge 1$.

\subsection{The general ${\cal N}=1$ supersymmetric action}

In this Section we consider a nonrelativistic particle of mass $m$ in an Euclidean space of dimension $d$,
whose trajectory is described by Cartesian coordinates $x_i(t)$ ($i=1,2,\ldots,d$). When coupled
to an arbitrary magnetic field $B_{ij}(x_i)$ (of which the normalisation includes the charge of the particle),
the action describing this system is given as\footnote{The implicit summation convention over repeated indices
is implied throughout, unless otherwise stated.} (in the absence of any other interaction),
\begin{equation}
S_0=\int dt\left(\frac{1}{2}m\dot{x}^2_i+\dot{x}_i A_i(x_i)\right),
\end{equation}
$A_i(x_i)$ being a vector potential from which the magnetic field derives through
\begin{equation}
B_{ij}(x_i)=\frac{\partial A_j(x_i)}{\partial x_i}\,-\,\frac{\partial A_i(x_i)}{\partial x_j},\qquad
B_{ji}(x_i)=-B_{ij}(x_i).
\end{equation}

In order to promote this dynamics without any supersymmetry, ${\cal N}=0$, to one
with a single supersymmetry, ${\cal N}=1$, let us extend the time variable $t$ into
a supertime space spanned not only by the real Grassmann even variable $t$ but also by
a single real Grassmann odd variable $\theta$, $\theta^*=\theta$, such that $\theta^2=0$.
Supertranslations in supertime are then generated by a supercharge of the form,
\begin{equation}
Q=\partial_\theta+i\theta\partial_t,\qquad Q^\dagger=Q,
\end{equation}
such that
\begin{equation}
Q^2=i\partial_t,
\end{equation}
all Grassmann odd derivatives being left derivatives.
One may also introduce a supercovariant derivative which anticommutes with the supercharge,
\begin{equation}
D=\partial_\theta - i\theta\partial_t,\qquad
D^\dagger=D,\qquad
\Big\{Q,D\Big\}=0,
\end{equation}
such that
\begin{equation}
D^2=-i\partial_t.
\end{equation}

Next, let us consider real Grassmann even supercoordinates $X_i(t,\theta)$, $X^*_i(t,\theta)=X_i(t,\theta)$,
of which the Grassmann expansion is necessarily of the form,
\begin{equation}
X_i(t,\theta)=x_i(t)+i\theta\lambda_i(t),\qquad i=1,2,\ldots,d.
\end{equation}
The variables $x_i(t)$ are Grassmann even, hence bosonic variables, corresponding to the real valued
coordinates of the particle in Euclidean space. The variables $\lambda_i(t)$ are Grassmann odd, hence fermionic
variables such that $\lambda^2_i=0$ (no summation over $i$), corresponding to real valued spin degrees of freedom for the
particle. Note that as $\theta$ is a scalar under Euclidean space transformations and in particular rotations,
both $x_i$ and $\lambda_i$ transform as SO($d$) vectors under spatial rotations.

Infinitesimal ${\cal N}=1$ supersymmetry transformations of these variables are generated by the action of $Q$ on these
supercoordinates,
\begin{equation}
\delta_\epsilon X_i(t,\theta)=-i\epsilon Q X_i(t,\theta),
\end{equation}
$\epsilon$ being a real Grassmann odd constant infinitesimal parameter. In component form one finds,
\begin{equation}
\delta_\epsilon x_i(t)=\epsilon\lambda_i(t),\qquad
\delta_\epsilon \lambda_i(t)=i\epsilon\dot{x}_i(t).
\label{eq:SUSY1}
\end{equation}

In a likewise manner, under any supersymmetry transformation the highest $\theta$ component of any quantity
defined over supertime space transforms as a total time derivative. Consequently, any action of the form
\begin{equation}
S=\int dt d\theta\,\mathbb{S}
\end{equation}
necessarily transforms by a total time derivative, hence is supersymmetry invariant. However since $S$ needs to be
Grassmann even, in the present instance of a single supersymmetry $\mathbb{S}$ needs to be Grassmann odd. Consequently,
$\mathbb{S}$ cannot be built out of $X_i$ alone, which excludes any interaction superpotential energy function of the
$x_i$, hence the $X_i$ alone, $W(X_i)$. The quantity $\mathbb{S}$ must involve derivatives of $X_i$.
In order to be consistent with the supersymmetry transformations, such derivatives must use the Grassmann odd
supercovariant derivative $D$, and then in odd multiples of it. However, the simplest choice of the form $DW(X_i)$
is of no use since upon the Grassmann integration the resulting contribution to the action proves to be a total time
derivative. Hence at least three factors of $D$ are required. Given the SO($d$) vector character of $X_i$ and
the requirement of a SO($d$) invariant kinetic energy contribution to the action, the next best choice is thus
$D^2X_i DX_i$, which indeed proves to be relevant,
\begin{equation}
\int dt d\theta \left\{D^2 X_i D X_i\right\}=\int dt\left\{-\dot{x}^2_i+i\dot{\lambda}_i\lambda_i\right\}.
\end{equation}
A similar reasoning yields for the coupling to the vector potential
\begin{equation}
\int dt d\theta\left\{DX_i A_i(X_i)\right\}=\int dt
\left\{-i\dot{x}_i A_i(x_i)-\frac{1}{2}B_{ij}(x_i)\lambda_i\lambda_j\right\}.
\end{equation}

In conclusion the action for the ${\cal N}=1$ supersymmetric nonrelativistic particle coupled to an arbitrary
background magnetic field in $d$ dimensional Euclidean space is given by,
\begin{eqnarray}
S_1 &=& \int dt d\theta\left\{-\frac{1}{2}mD^2 X_i DX_i + i DX_i A_i(X_i)\right\} \nonumber \\
&=& \int dt\left\{\frac{1}{2}m\dot{x}^2_i-\frac{1}{2}im\dot{\lambda}_i\lambda_i + \dot{x}_i A_i(x_i)
-\frac{1}{2}i B_{ij}(x_i)\lambda_i\lambda_j\right\}.
\end{eqnarray}
By construction, this action is invariant under the infinitesimal supersymmetry transformations (\ref{eq:SUSY1}),
as may also be checked by explicit inspection using the last expression for $S$ in terms of the component coordinates
and their supersymmetry transformations. What is noteworthy about this
system is that ${\cal N}=1$ supersymmetry precludes any possible interaction potential energy besides the
magnetic coupling. Thus for instance an electrostatic or electric coupling of such a charged particle is
incompatible with ${\cal N}=1$ supersymmetry.

The conserved Noether generator for supersymmetry transformations is readily found to be given by
\begin{equation}
Q_{\rm Noether}=im\lambda_i\dot{x}_i.
\end{equation}
Unless the background magnetic field display specific symmetry properties, the system does not possess any
further conserved quantities besides $Q_{\rm Noether}$ and its total energy, namely its canonical Hamiltonian.

The Hamiltonian formulation of the system is provided by conjugate pairs of Grassmann even and Grassmann odd phase
space variables. The bosonic coordinates, $x_i$, possess Grassmann even conjugate momenta, $p_i=m\dot{x}_i+A_i(x_i)$, with the
canonical Poisson brackets,
\begin{equation}
\Big\{x_i,p_j\Big\}=\delta_{ij}.
\end{equation}
The action being already first-order in $\dot{\lambda}_i$, the fermionic variables $\lambda_i$ are
conjugate to themselves, with the following Grassmann graded Poisson brackets\footnote{Applying Dirac's analysis of constraints,
these brackets are the Dirac brackets resulting from fermionic second-class constraints\cite{govaerts}.},
\begin{equation}
\left\{\lambda_i,\lambda_j\right\}=-i\frac{1}{m}\delta_{ij}.
\end{equation}
The canonical Hamiltonian then reads,
\begin{equation}
H=\frac{1}{2m}\Big(p_i-A_i(x_i)\Big)^2+\frac{1}{2}i B_{ij}\lambda_i\lambda_j,
\end{equation}
while for the Noether supercharge,
\begin{equation}
Q_{\rm Noether}=i\lambda_i\Big(p_i-A_i(x_i)\Big).
\end{equation}

Quantisation of the system is straightforward enough. Upon rescaling of the fermionic operators an anticommuting
Clifford algebra is obtained in the Grassmann odd sector, while the Grassmann even one obeys the usual Heisenberg
algebra of Euclidean space. Namely, writing
\begin{equation}
\lambda_i=\sqrt{\frac{\hbar}{2m}}\gamma_i,\qquad \lambda^\dagger_i=\lambda_i,\qquad \gamma^\dagger_i=\gamma_i,
\end{equation}
the quantum system is defined by the algebra of the following (only nonvanishing) commutation and anticommutation relations of
hermitian operators,
\begin{equation}
\Big[x_i,p_j\Big]=i\hbar\delta_{ij}\mathbb{I},\qquad
\Big\{\gamma_i,\gamma_j\Big\}=2\delta_{ij}\mathbb{I},
\end{equation}
while one has,
\begin{equation}
H=\frac{1}{2m}\Big(p_i-A_i(x_i)\Big)^2+\frac{\hbar}{m}B_{ij}(x_i)\sigma_{ij},\qquad
Q_{\rm Noether}=i\sqrt{\frac{\hbar}{2m}}\gamma_i\Big(p_i-A_i(x_i)\Big),
\end{equation}
where
\begin{equation}
\sigma_{ij}=\frac{1}{2}i\Big[\gamma_i,\gamma_j\Big],
\end{equation}
generate the spinor representation of SO($d$). Hence indeed the system describes a spin $1/2$ charged point particle
coupled to the magnetic field, possessing a total energy which includes the magnetic spin coupling energy for a
gyromagnetic factor of $g=2$. This very specific value for this coupling is a direct consequence of supersymmetry,
as is well known.

An explicit analytic solution of the quantum system is possible only for specific magnetic field configurations. A constant field is certainly such a case, with the vector
potential then linear in the coordinates for a particular class of gauges. Through an appropriate rotation
in Euclidean space it is always possible to bring the magnetic field $B_{ij}$ to lie only in the $(12)$ plane,
say, in which case the motion of the particle in all directions perpendicular to that plane is free,
any additional potential being forbidden by the ${\cal N}=1$ supersymmetry.
Consequently, let us henceforth restrict the discussion to the ${\cal N}=1$ supersymmetric Landau problem
in the plane.

\subsection{Landau levels and ${\cal N}=1$ supersymmetry}

Restricting to the Landau problem in the plane, the magnetic field $B_{12}$ of the previous Section
corresponds to that of Section~\ref{Sect2}, $B=B_{12}$. Working once again in the circular gauge for the vector potential,
the action of the system is manifestly invariant under SO(2) rotations in the plane, $i,j=1,2$.
The Noether generator for such infinitesimal rotations is found to be,
\begin{equation}
L_{\rm Noether}=\epsilon_{ij}x_i\Big(m\dot{x}_j+A_j(x_i)\Big)-\frac{1}{2}im\epsilon_{ij}\lambda_i\lambda_j=
x_1 p_2 - x_2 p_1 - \frac{1}{2}im\left[\lambda_1,\lambda_2\right],
\end{equation}
$\epsilon_{ij}$ being the antisymmetric symbol with $\epsilon_{12}=+1$. This quantity thus defines the total
angular-momentum of the system, inclusive of the particle spin contribution.

Hence, given this magnetic field configuration, the system possesses three conserved quantities, of which the Poisson brackets
all vanish on account of their invariance under supersymmetry and translations in time, namely $H$, $Q_{\rm Noether}$ and $L_{\rm Noether}$.
At the quantum level, these three operators thus commute and may be diagonalised in a common basis of eigenstates.

Note also that the Grassmann graded phase space of the system consists of 4 bosonic real variables, $(x_i,p_i)$, and
2 fermionic real variables, $\lambda_i$. Phase space is thus of dimension $(4|2)$ in that sense.

In the present case $d=2$ and the operators $\gamma_i=\lambda_i\sqrt{2m/\hbar}$ define the SO(2) Clifford algebra,
$\left\{\gamma_i,\gamma_j\right\}=2\delta_{ij}\mathbb{I}$.  A possible representation\footnote{Any other representation is unitarily equivalent
to the present one by some SU(2) transformation acting on the Pauli matrices.} of the fermionic sector of the system is thus given by the Pauli matrices $\sigma_\alpha$ ($\alpha=1,2,3$) as follows
\begin{equation}
\gamma_1=\sigma_1,\qquad
\gamma_2=\sigma_2,\qquad
\Big[\gamma_1,\gamma_2\Big]=2i\sigma_3.
\end{equation}
Consequently one has
\begin{equation}
H=\frac{1}{2m}\left(p_1+\frac{1}{2}Bx_2\right)^2+\frac{1}{2m}\left(p_2-\frac{1}{2}Bx_1\right)^2
-\frac{\hbar}{2m}B\sigma_3,
\end{equation}
\begin{equation}
L_{\rm Noether}=x_1 p_2 - x_2 p_1 +\frac{1}{2}\hbar\sigma_3,
\end{equation}
\begin{equation}
Q_{\rm Noether}=i\sqrt{\frac{\hbar}{2m}}\left[\sigma_1\left(p_1+\frac{1}{2}Bx_2\right)\,+\,
\sigma_2\left(p_2-\frac{1}{2}Bx_1\right)\right],
\end{equation}
making it explicit that indeed the particle is of spin $1/2$, with an energy which is decreased when the spin projection
onto the axis perpendicular to the plane is aligned with the magnetic field, as it should. Spin up and down states differ
in energy by the quantum of energy $\hbar\omega_c$, which is also the quantum of energy excitations in the bosonic sector.
This equality of energy quanta values in both sectors is required by supersymmetry, and is intimately related
to the value $g=2$ for the gyromagnetic factor of the charged particle. Incidentally, this equality of bosonic and fermionic
energy gaps confirms once again that ${\cal N}=1$ supersymmetry forbids any extra potential, which would
otherwise break the equal spacing in energy of Landau levels and lift their degeneracies in an {\it a priori\/} arbitrary fashion
in the bosonic sector.

Using in the bosonic sector the chiral Fock algebras defined in Section~\ref{Sect2}, and introducing the following
chiral combinations in the fermionic spin 1/2 sector as well,
\begin{equation}
\sigma_\pm=\frac{1}{2}\left(\sigma_1 \pm i\sigma_2\right),\qquad
\sigma^\dagger_\pm=\sigma_\mp,\qquad
\sigma_1=\sigma_+ + \sigma_-,\qquad
\sigma_2=-i\left(\sigma_+ - \sigma_-\right),
\end{equation}
the algebra of the elementary degrees of freedom of the system is,
\begin{equation}
\Big[a_\pm,a^\dagger_\pm\Big]=\mathbb{I},\qquad
\Big\{\sigma_+,\sigma_-\Big\}=\mathbb{I},\qquad
\sigma^2_+=0,\qquad \sigma^2_-=0.
\end{equation}
Furthermore, one also has,
\begin{equation}
\Big\{\sigma_3,\sigma_\pm\Big\}=\pm 2\sigma_\pm,\qquad
\Big[\sigma_+,\sigma_-\Big]=\sigma_3,\qquad
\Big[\sigma_3,\sigma_\pm\Big]=\pm 2\sigma_\pm.
\end{equation}

By direct substitution one finds,
\begin{equation}
H=\hbar\omega_c\left(a^\dagger_- a_- + \frac{1}{2}\right)-\frac{1}{2}\hbar\omega_c\sigma_3=
\hbar\omega_c\left(a^\dagger_- a_- +\frac{1-\sigma_3}{2}\right),
\end{equation}
\begin{equation}
L\equiv L_{\rm Noether}=\hbar\left(a^\dagger_+ a_+ - a^\dagger_- a_-+\frac{1}{2}\sigma_3\right),
\end{equation}
\begin{equation}
Q_0=\frac{i}{\sqrt{\hbar}}Q_{\rm Noether}=i\sqrt{\hbar\omega_c}\left(\sigma_- a_-
\,-\,\sigma_+ a^\dagger_-\right).
\end{equation}
The normalisation and phase of $Q_0$ are chosen such that
\begin{equation}
Q^2_0=H,\qquad Q^\dagger_0=Q_0.
\end{equation}
The action of these operators on the degrees of freedom is,
\begin{equation}
\Big[H,a^{(\dagger)}_+\Big]=0,\qquad
\Big[H,a^\dagger_-\Big]=\hbar\omega_c\,a^\dagger_-,\qquad
\Big[H,a_-\Big]=-\hbar\omega_c\,a_-,\qquad
\Big[H,\sigma_\pm\Big]=\mp\hbar\omega_c\,\sigma_\pm,
\end{equation}
\begin{equation}
\Big[L,a^\dagger_\pm\Big]=\pm\hbar\,a^\dagger_\pm,\qquad
\Big[L,a_\pm\Big]=\mp\hbar\,a_\pm,\qquad
\Big[L,\sigma_\pm\Big]=\pm\hbar\,\sigma_\pm,
\end{equation}
\begin{equation}
\Big[Q_0,a^{(\dagger)}_+\Big] = 0,\qquad
\Big[Q_0,a^\dagger_-\Big]=i\sqrt{\hbar\omega_c}\,\sigma_-,\qquad
\Big[Q_0,a_-\Big]=i\sqrt{\hbar\omega_c}\,\sigma_+,
\end{equation}
\begin{equation}
\Big\{Q_0,\sigma_+\Big\} = i\sqrt{\hbar\omega_c}\,a_-,\qquad
\Big\{Q_0,\sigma_-\Big\}=-i\sqrt{\hbar\omega_c}\,a^\dagger_-.
\end{equation}

It may easily be checked that all commutators of these three operators do indeed vanish,
\begin{equation}
\Big[L,H\Big]=0,\qquad
\Big[Q_0,H\Big]=0,\qquad
\Big[Q_0,L\Big]=0.
\end{equation}
Note that ${\cal N}=1$ supersymmetry transformations mix only the left-handed chiral bosonic mode with the fermionic
degrees of freedom, leaving the right-handed chiral bosonic mode invariant. This simplest form of supersymmetry thus
has a ``chiral preference" in the presence of the magnetic field which distinguishes these two chiralities and breaks
time reversal invariance, in spite of the fact that supersymmetry transformations treat spin up and down states both
on an equal footing. Incidentally, this property of ${\cal N}=1$ supersymmetry transformations implies that the specific
combinations $(p_1+Bx_2/2)$ and $(p_2-Bx_1/2)$ in (\ref{eq:pair2}) are supersymmetry invariants.

A construction of a basis of states for the quantised system, which furthermore diagonalises these operators,
is obvious. In the bosonic sector one has the chiral Fock states $|n_+,n_-\rangle$ of Section~\ref{Sect2},
while in the fermionic sector one has the spin up and spin down states, $|s=\pm 1\rangle$, of the
two dimensional Hilbert space spanning the Clifford algebra of $\gamma_i$, which are eigenstates of
the spin operator $\sigma_3$,
\begin{equation}
\sigma_3|s\rangle=s|s\rangle,\qquad
\langle s|s'\rangle=\delta_{s,s'},\qquad s,s'=+1,-1.
\end{equation}
The full Hilbert space is obtained as the tensor product of these two representation spaces,
leading to the orthonormalised basis $|n_+,n_-;s\rangle$ such that,
\begin{equation}
\langle n_+,n_-;s|m_+,m_-;s'\rangle=\delta_{n_+,m_+}\delta_{n_-,m_-}\delta_{s,s'},\qquad
\sum_{n_+,n_-=0}^\infty\sum_{s=\pm 1}|n_+,n_-;s\rangle\langle n_+,n_-;s|=\mathbb{I}.
\end{equation}

These states diagonalise the Hamiltonian and angular-momentum operators already, but not yet the
supercharge $Q_0$,
\begin{eqnarray}
H|n_+,n_-;s\rangle &=& \hbar\omega_c\left(n_-+\frac{1-s}{2}\right)|n_+,n_-;s\rangle, \nonumber \\
L|n_+,n_-;s\rangle &=& \hbar\left(n_+ - n_- +\frac{1}{2}s\right)|n_+,n_-;s\rangle .
\end{eqnarray}
Note again the large degeneracy in the Landau levels, associated to the right-handed excitations of the
bosonic sector, but extended further in presence of the ${\cal N}=1$ supersymmetry by a degeneracy in
opposite spin values, $s=-1,+1$, for two adjacent bosonic Landau levels in $n_-=N-1,N$, respectively.
For any fixed value of $N=1,2,\ldots$ and whatever values $n_+,m_+=0,1,2,\ldots$, the states
$|n_+,N-1;s=-1\rangle$ and $|m_+,N;s=+1\rangle$ are degenerate with energy $\hbar\omega_c N$.

This remark does not apply to the lowest energy sector, or lowest Landau level of the system,
which consists of all the states $|n_+,0;s=+1\rangle$ of vanishing energy ($n_+=0,1,2,\ldots$),
and which are in fact supersymmetry invariant,
\begin{equation}
Q_0|n_+,0;s=+1\rangle=0.
\end{equation}
The lowest Landau level thus provides a trivial representation of ${\cal N}=1$ supersymmetry, each of
its states being supersymmetry invariant. This also implies that supersymmetry remains unbroken by the
quantum dynamics of the system.

For all the other Landau levels one has a non trivial ${\cal N}=1$ supersymmetry
transformation and, as a matter of fact, a two dimensional reducible representation for any given $n_+=0,1,2,\ldots$
and $N\ge 1$,
\begin{eqnarray}
Q_0|n_+,N;s=+1\rangle &=& i\sqrt{\hbar\omega N}\,|n_+,N-1;s=-1\rangle, \nonumber \\
Q_0|n_+,N-1;s=-1\rangle &=& -i\sqrt{\hbar\omega_c N}\,|n_+,N;s=+1\rangle.
\end{eqnarray}
This result is obviously consistent with the property $Q^2_0=H$ since these states
are already each an eigenstate of $H$ with energy $\hbar\omega_c N$. Furthermore, note each is also an eigenstate
of the angular-momentum operator $L$ with common eigenvalue, $\hbar(n_+-N+1/2)$.
Consequently, for any given values for $n_+=0,1,2,\ldots$ and $N\ge 1$, one has the following two orthonormalised
eigenstates of the ${\cal N}=1$ supercharge $Q_0$,
\begin{equation}
|n_+,N;\delta\rangle\equiv\frac{1}{\sqrt{2}}\Big[\,
|n_+,N;s=+1\rangle\,+\,i\delta\,|n_+,N-1;s=-1\rangle\,\Big],\qquad \delta=\pm 1,
\end{equation}
such that
\begin{equation}
Q_0|n_+,N;\delta\rangle = \delta\sqrt{\hbar\omega_c N}\,|n_+,N;\delta\rangle,\qquad
N\ge 1,\quad \delta=\pm 1.
\end{equation}
These eigenvalues are real since $Q_0$ is hermitian, and correspond to the two square roots of the energy
eigenvalue of that Landau level, $\hbar\omega_c N$.

In conclusion, an orthonormalised basis\footnote{Up to arbitrary phase redefinitions of each state, this basis is unique.},
which diagonalises all three commuting operators $H$, $L$ and $Q_0$, consists of the sets of states
$|n_+,0;s=+1\rangle$ and $|n_+,N;\delta\rangle$
with $n_+=0,1,2,\ldots$, $N=1,2,\ldots$, $\delta=\pm 1$ and the following resolution of the unit operator,
\begin{equation}
\mathbb{I}=
\sum_{n_+=0}^\infty|n_+,0;s=+1\rangle\langle n_+,0;s=+1|\,+\,
\sum_{\delta=\pm 1}\sum_{N=1}^\infty\sum_{n_+=0}^\infty|n_+,N;\delta\rangle\langle n_+,N;\delta|.
\end{equation}
Furthermore
\begin{eqnarray}
H|n_+,0;s=+1\rangle &=& 0,\qquad
L|n_+,0;s=+1\rangle=\hbar\left(n_++\frac{1}{2}\right)|n_+,0;s=+1\rangle, \nonumber \\
Q_0|n_+,0;s=+1\rangle &=&0,
\end{eqnarray}
while for $N\ge 1$, $\delta=\pm 1$,
\begin{eqnarray}
H|n_+,N;\delta\rangle &=& \hbar\omega_c N |n_+,N;\delta\rangle,\qquad
L|n_+,N;\delta\rangle=\hbar\left(n_+-N+\frac{1}{2}\right)|n_+,N;\delta\rangle, \nonumber \\
Q_0|n_+,N;\delta\rangle &=& \delta\sqrt{\hbar\omega_c N}\,|n_+,N;\delta\rangle.
\end{eqnarray}

For the discussion that follows, it is also useful to have available spectral decompositions
of all elementary degrees of freedom in this basis. Let us introduce the following
notations for the projection operators ($N\ge 1$)
\begin{equation}
\mathbb{P}_0=\sum_{n_+=0}^\infty |n_+,0;s=+1\rangle\langle n_+,0;s=+1|,\qquad
\mathbb{P}(N,\delta)=\sum_{n_+=0}^\infty|n_+,N;\delta\rangle\langle n_+,N;\delta|,
\end{equation}
which are such that,
\begin{equation}
\mathbb{I}=\mathbb{P}_0+\sum_{\delta=\pm 1}\sum_{N=1}^\infty\mathbb{P}(N,\delta),\quad
H=\hbar\omega_c\sum_{\delta=\pm 1}\sum_{N=1}^\infty N\mathbb{P}(N,\delta),\quad
Q_0=\sqrt{\hbar\omega_c}\sum_{\delta=\pm 1}\sum_{N=1}^\infty \delta\mathbb{P}(N,\delta).
\end{equation}
It is also convenient to introduce the following two quantities,
\begin{equation}
F_+(N)=\frac{\sqrt{N}+\sqrt{N-1}}{2},\qquad
F_-(N)=\frac{\sqrt{N}-\sqrt{N-1}}{2},\qquad N\ge 1,
\end{equation}
in terms of which one finds the following representations for the elementary degrees of freedom
\begin{eqnarray}
a_+ &=& \sum_{n_+=0}^\infty|n_+,0;s=+1\rangle\,\sqrt{n_+ + 1}\,\langle n_+ + 1,0;s=+1|\,+\, \nonumber \\
&& + \sum_{\delta=\pm 1}\sum_{N=1}^\infty\sum_{n_+=0}^\infty |n_+,N;\delta\rangle\,\sqrt{n_+ +1}\,
\langle n_+ + 1,N;\delta|, \\
a^\dagger_+ &=& \sum_{n_+=0}^\infty|n_+ +1,0;s=+1\rangle\,\sqrt{n_++1}\,\langle n_+,0;s=+1|\,+\, \nonumber \\
&& +\sum_{\delta=\pm 1}\sum_{N=1}^\infty\sum_{n_+=0}^\infty|n_+ +1,N;\delta\rangle\,\sqrt{n_+ + 1}\,
\langle n_+,N;\delta|,
\end{eqnarray}
\begin{eqnarray}
a_- &=& \sum_{\delta=\pm 1}\sum_{n_+=0}^\infty |n_+,0;s=+1\rangle\,\frac{1}{\sqrt{2}}\,\langle n_+,1;\delta|\,+\, \nonumber \\
&& +\sum_{\delta=\pm 1}\sum_{N=1}^\infty\sum_{n_+=0}^\infty |n_+,N;\delta\rangle\,F_+(N)\,\langle n_+,N+1;\delta|\,+\, \nonumber \\
&& +\sum_{\delta=\pm 1}\sum_{N=1}^\infty\sum_{n_+=0}^\infty |n_+,N;\delta\rangle\,F_-(N)\,\langle n_+,N+1;-\delta|, \\
a^\dagger_- &=& \sum_{\delta=\pm 1}\sum_{n_+=0}^\infty |n_+,1;\delta\rangle\,\frac{1}{\sqrt{2}}\,\langle n_+,0;s=+1|\,+\, \nonumber \\
&& +\sum_{\delta=\pm 1}\sum_{N=1}^\infty\sum_{n_+=0}^\infty |n_+,N+1;\delta\rangle\,F_+(N)\,\langle n_+,N;\delta|\,+\, \nonumber \\
&& +\sum_{\delta=\pm 1}\sum_{N=1}^\infty\sum_{n_+=0}^\infty |n_+,N+1;\delta\rangle\,F_-(N)\,\langle n_+,N;-\delta|,
\end{eqnarray}
and,
\begin{eqnarray}
\sigma_+ &=& \sum_{\delta=\pm 1}\sum_{n_+=0}^\infty |n_+,0;s=+1\rangle\,\frac{i\delta}{\sqrt{2}}\,\langle n_+,1;\delta|\,+\, \nonumber \\
&& +\sum_{\delta,\delta'=\pm 1}\sum_{N=1}^\infty\sum_{n_+=0}^\infty
|n_+,N;\delta\rangle\,\frac{i\delta'}{2}\,\langle n_+,N+1;\delta'|,\\
\sigma_- &=& \sum_{\delta=\pm 1}\sum_{n_+=0}^\infty |n_+,1;\delta\rangle\,\frac{-i\delta}{\sqrt{2}}\,\langle n_+,0;s=+1|\,+\, \nonumber \\
&& +\sum_{\delta,\delta'=\pm 1}\sum_{N=1}^\infty\sum_{n_+=0}^\infty
|n_+,N+1;\delta\rangle\,\frac{-i\delta}{2}\,\langle n_+,N;\delta'|.
\end{eqnarray}

\section{Supersymmetric Landau Level Projections}

In the same spirit that led to the noncommutative Moyal--Voros plane by projection of the ordinary ${\cal N}=0$ Landau problem onto any of its Landau levels, we would now like to consider similar projections which preserve
the ${\cal N}=1$ supersymmetry in order to identify a ${\cal N}=1$ supersymmetric extended
non(anti)commutative Moyal--Voros superplane.

Clearly the supersymmetric invariant lowest Landau level
$|n_+,0;s=+1\rangle$ is of no use in that respect. A projection onto that single level produces the
ordinary ${\cal N}=0$ noncommutative Moyal--Voros plane of Section~\ref{Sect2}. Likewise, as may also be seen from the above expressions,
projecting onto a single supersymmetric covariant Landau level $|n_+,N;\delta\rangle$ ($N\ge 1$), using $\mathbb{P}(N,\delta)$,
once again results in the projection of the spin degrees of freedom, $\sigma_\pm$, and the left-handed chiral
bosonic modes, $(a_-,a^{\dagger}_-)$, onto the null operator, leaving simply the bosonic
noncommutative Moyal--Voros plane of Section~\ref{Sect2} spanned by $(a_+,a^\dagger_+)$. One thus has to resort to more
than a single Landau sector $|n_+,N;\delta\rangle$ in order to gain something new.

Restricting to a single energy eigenvalue, $\hbar\omega_c N$ with $N\ge 1$ say, a natural choice appears to be given by a projector combining the two supersymmetry eigenvalues with $\delta=\pm 1$,
\begin{equation}
\mathbb{P}(N)=\mathbb{P}(N,\delta=+1)+\mathbb{P}(N,\delta=-1).
\end{equation}
This projection is such that (again in the notation that $\overline{A}=\mathbb{P}(N) A\,\mathbb{P}(N)$ for some operator $A$),
\begin{equation}
\overline{\sigma}_+=0,\qquad
\overline{\sigma}_-=0,\qquad
\overline{a}_-=0,\qquad
\overline{a^\dagger_-}=\overline{a}^\dagger_-=0,
\end{equation}
while
\begin{eqnarray}
\overline{a}_+ &=& \sum_{\delta=\pm 1}\sum_{n_+=0}^\infty |n_+,N;\delta\rangle\,\sqrt{n_+ + 1}\,\langle n_+ +1,N;\delta|, \nonumber \\
\overline{a^\dagger_+}=\overline{a}^\dagger_+ &=&
\sum_{\delta=\pm 1}\sum_{n_+=0}^\infty |n_+ +1,N;\delta\rangle\,\sqrt{n_+ + 1}\,\langle n_+,N;\delta|,
\end{eqnarray}
as well as,
\begin{eqnarray}
\overline{\sigma}_3 &=& \mathbb{P}(N),\qquad
\overline{H} = \hbar\omega_c N \mathbb{P}(N),\qquad\qquad
\overline{Q}_0 = \sqrt{\hbar\omega_c}\Big[\mathbb{P}(N,+1)-\mathbb{P}(N,-1)\Big], \nonumber \\
\overline{L} &=& \hbar\sum_{\delta=\pm 1}\sum_{n_+=0}^\infty |n_+,N;\delta\rangle\,\left(n_+-N+\frac{1}{2}\right)\,
\langle n_+,N;\delta|.
\end{eqnarray}

Hence, using the projector $\mathbb{P}(N)$ one in fact simply recovers two commuting copies of the ${\cal N}=0$ noncommutative
Moyal--Voros plane, distinguished by the sign $\delta=\pm 1$ of the $Q_0$ eigenvalues at Landau level $N$. Both the projected
left-handed chiral bosonic modes, $(a_-,a^\dagger_-)$ (as in Section~\ref{Sect2}), and the fermionic modes, $\sigma_\pm$,
are reduced to the null operator, leaving only the Fock algebra of the projected right-handed chiral bosonic modes,
\begin{equation}
\left[\overline{a}_+,\overline{a}^\dagger_+\right]=\mathbb{P}(N).
\end{equation}
Consequently, in order to gain some new structure, one needs to consider a projection involving at least two\footnote{Like in the
${\cal N}=0$ case, including more than two energy levels does not seem to offer any particular advantages.} Landau sectors
separated by the quantum energy gap $\hbar\omega_c$.

\subsection{The two Landau level projection}

Given a pair of fixed values $\delta=\pm 1$ and $N\ge 1$, let us consider the projection associated
to the projector $\mathbb{P}=\mathbb{P}_+$ defined as, together with its counterpart $\mathbb{P}_-$,
\begin{equation}
\mathbb{P}=\mathbb{P}_+=\mathbb{P}(N,\delta)+\mathbb{P}(N+1,\epsilon\delta),\qquad
\mathbb{P}_-=\mathbb{P}(N,\delta)-\mathbb{P}(N+1,\epsilon\delta),
\end{equation}
where $\epsilon=\pm 1$. In this manner the projection onto the two energy sectors at levels $N$ and $(N+1)$
involves either the same or opposite signs for the supercharge eigenvalues. Including all possibilities $\delta=\pm 1$
and $\epsilon=\pm 1$ accounts for all four such combinations given the levels $N$ and $(N+1)$.

This projection is such that,
\begin{equation}
\overline{\sigma}_3=0,
\end{equation}
while for the fermionic degrees of freedom,
\begin{equation}
\overline{\sigma}_+=\sum_{n_+=0}^\infty |n_+,N;\delta\rangle\,\frac{i\epsilon\delta}{2}\,\langle n_+,N+1;\epsilon\delta|,\qquad
\overline{\sigma}_-=\sum_{n_+=0}^\infty |n_+,N+1;\epsilon\delta\rangle\,\frac{-i\epsilon\delta}{2}\,\langle n_+,N;\delta|,
\end{equation}
which are such that,
\begin{equation}
\overline{\sigma}^\dagger_+=\overline{\sigma}_-,\qquad
\overline{\sigma}^\dagger_-=\overline{\sigma}_+.
\end{equation}
The projected bosonic degrees of freedom are given as,
\begin{eqnarray}
\overline{a}_- &=& \sum_{n_+=0}^\infty |n_+,N;\delta\rangle\,F_\epsilon(N)\,\langle n_+,N+1;\epsilon\delta|, \nonumber \\
\overline{a^\dagger_-} = \overline{a}^\dagger_- &=& \sum_{n_+=0}^\infty
|n_+,N+1;\epsilon\delta\rangle\,F_\epsilon(N)\,\langle n_+,N;\delta|,
\end{eqnarray}
and,
\begin{eqnarray}
\overline{a}_+ &=& \ \ \ \sum_{n_+=0}^\infty |n_+,N;\delta\rangle\,\sqrt{n_+ +1}\,\langle n_+ +1,N;\delta|\,+\, \nonumber \\
 && +\sum_{n_+=0}^\infty |n_+,N+1;\epsilon\delta\rangle\,\sqrt{n_+ +1}\,\langle n_+ +1,N+1;\epsilon\delta| , \nonumber \\
\overline{a^\dagger_+}=\overline{a}^\dagger_+ &=&
\ \ \ \sum_{n_+=0}^\infty |n_+ +1,N;\delta\rangle\,\sqrt{n_+ +1}\,\langle n_+,N;\delta| \,+\, \nonumber \\
 && +\sum_{n_+=0}^\infty |n_+ +1,N+1;\epsilon\delta\rangle\,\sqrt{n_+ +1}\,\langle n_+,N+1;\epsilon\delta|.
\end{eqnarray}
For the remaining operators of interest, one has,
\begin{eqnarray}
\overline{H} &=& \hbar\omega_c\Big[ N\mathbb{P}(N,\delta) + (N+1)\mathbb{P}(N+1,\epsilon\delta)\Big]=\overline{Q}^2_0, \nonumber \\
\overline{Q}_0 &=& \delta\sqrt{\hbar\omega_c}\Big[ \sqrt{N}\mathbb{P}(N,\delta) +\epsilon \sqrt{N+1}\mathbb{P}(N+1,\epsilon\delta)\Big], \nonumber \\
\overline{L} &=& \hbar\left[\overline{a}^\dagger_+\overline{a}_+-\left(N-\frac{1}{2}\right)\mathbb{P}(N,\delta)
-\left(N+\frac{1}{2}\right)\mathbb{P}(N+1,\epsilon\delta)\right].
\end{eqnarray}
Hence this projection certainly leads to some new algebraic structure consistent with the ${\cal N}=1$ supersymmetry
since we still have $\overline{Q}^2_0=\overline{H}$, while the projected left-handed chiral bosonic as well
as the fermionic contents remain non trivial. Yet, like in the ${\cal N}=0$ case the number of projected phase space degrees of freedom is reduced by two since the latter two sets of degrees of freedom are no longer independent.
Indeed, a direct inspection of the above expressions shows that one has
\begin{equation}
\overline{a}_- + 2i\epsilon\delta F_\epsilon(N)\,\overline{\sigma}_+=0,\qquad
\overline{a}^\dagger_- - 2i\epsilon\delta F_\epsilon(N)\,\overline{\sigma}_-=0.
\label{eq:iden1}
\end{equation}
In other words, rather than the two combinations $(p_1+Bx_2/2)$ and $(p_2-Bx_1/2)$, which upon projection vanish
in the ${\cal N}=0$ case\footnote{These two combinations do not vanish for the present ${\cal N}=1$ 
supersymmetry covariant projection.}, the two combinations of degrees of freedom that vanish given the present projection in
the ${\cal N}=1$ case are those above.

This projection thus effects a boson-fermion transmutation of sorts! Bosonic degrees of freedom $\overline{a}^{(\dagger)}_-$
obey specific commutation relations, while fermionic degrees of freedom $\overline{\sigma}_\pm$ obey specific anticommutation
relations, which in each case specify precisely this spin-statistics property. However, through the considered projection
by $\mathbb{P}=\mathbb{P}_+$, these two sets of degrees of freedom become identified and coalesce into one another, and
hence are characterised by both commutation and anticommutation relations which are independent from, but consistent with one another.
The considered ${\cal N}=1$ supersymmetric invariant projection results in two bosonic and two fermionic phase space variables to become
identified, and yet preserving both these statistics properties. Since this proves to be convenient, hereafter we
express quantities in terms of $\overline{\sigma}_\pm$ only, knowing that at the same time these variables stand for
$\overline{a}^{(\dagger)}_-$ as well. Given this dual r\^ole, the variables $\overline{\sigma}_\pm$ are thus characterised
by both commutation and anticommutation properties. For instance, besides those operators already mentioned above,
the projected supercharge $\overline{Q}_0$ also possesses well defined commutation and anticommutation relations
with these bosonic-fermionic variables. This is quite an intriguing outcome of the effected projection.

\subsection{The algebra of the ${\cal N}=1$ supersymmetric Moyal--Voros superplane}

Since the commutators and anticommutators of $\overline{\sigma}_\pm$ are now both specified, so are their bilinear products.
One finds,
\begin{equation}
\overline{\sigma}^2_+=0,\qquad
\overline{\sigma}^2_-=0,\qquad
\overline{\sigma}_+\overline{\sigma}_-=\frac{1}{4}\mathbb{P}(N,\delta),\qquad
\overline{\sigma}_-\overline{\sigma}_+=\frac{1}{4}\mathbb{P}(N+1,\epsilon\delta).
\end{equation}
Introducing the operators
\begin{equation}
\overline{\sigma}_1=\overline{\sigma}_+ + \overline{\sigma}_-,\qquad
\overline{\sigma}_2=-i\left(\overline{\sigma}_+ - \overline{\sigma}_-\right),
\end{equation}
one has likewise,
\begin{equation}
\overline{\sigma}^2_1=\frac{1}{4}\mathbb{P}_+,\qquad
\overline{\sigma}^2_2=\frac{1}{4}\mathbb{P}_+,\qquad
\overline{\sigma}_1\overline{\sigma}_2=\frac{1}{4}i\mathbb{P}_-,\qquad
\overline{\sigma}_2\overline{\sigma}_1=-\frac{1}{4}i\mathbb{P}_-.
\end{equation}
Consequently, the full algebra of commutation and anticommutation relations for these bosonic-fermionic
variables is given as,
\begin{equation}
\overline{\sigma}^2_+=0,\qquad
\overline{\sigma}^2_-=0,\qquad
\Big\{\overline{\sigma}_+,\overline{\sigma}_-\Big\}=\frac{1}{4}\mathbb{P}_+,\qquad
\Big[\overline{\sigma}_+,\overline{\sigma}_-\Big]=\frac{1}{4}\mathbb{P}_-,
\end{equation}
\begin{equation}
\Big\{\overline{\sigma}_i,\overline{\sigma}_j\Big\}=\frac{1}{4}\,2\delta_{ij}\,\mathbb{P}_+,\qquad
\Big[\overline{\sigma}_i,\overline{\sigma}_j\Big]=\frac{1}{2}i\epsilon_{ij}\,\mathbb{P}_-.
\label{eq:fermion}
\end{equation}
Note the consistency of these expressions with the manifest SO(2) covariance properties under rotations in the plane.

Besides these two bosonic-fermionic variables, the ${\cal N}=1$ Moyal--Voros superplane also consists of the projected
right-handed chiral bosonic variables $a^{(\dagger)}_+$, making up a total of four variables for that non(anti)commutative
variety. This sector commutes with the previous one,
\begin{equation}
\Big[\overline{a}_+,\overline{\sigma}_\pm\Big]=0,\qquad
\Big[\overline{a}^\dagger_+,\overline{\sigma}_\pm\Big]=0,\qquad
\Big[\overline{a}_+,\overline{\sigma}_i\Big]=0,\qquad
\Big[\overline{a}^\dagger_+,\overline{\sigma}_i\Big]=0,
\end{equation}
while these two operators also define a Fock algebra on the projected space,
\begin{equation}
\Big[\overline{a}_+,\overline{a}^\dagger_+\Big]=\mathbb{P}_+.
\end{equation}
In effect this sector alone consists of two copies of the ${\cal N}=0$ Moyal--Voros plane, soldered with the remaining sector of bosonic-fermionic variables $\overline{\sigma}_\pm$,
to build up the non(anti)com\-mu\-ta\-ti\-ve space of the projected original Cartesian coordinates of the
Euclidean plane, as we now proceed to show.

Using the relations in (\ref{eq:invert}), one finds for the projected Cartesian coordinates,
\begin{equation}
\overline{x}_1=\sqrt{\frac{\hbar}{2B}}\Big[\left(\overline{a}_+ + \overline{a}^\dagger_+\right)
+ 2\epsilon\delta\, F_\epsilon(N)\, \overline{\sigma}_2\Big],\qquad
\overline{x}_2=\sqrt{\frac{\hbar}{2B}}\Big[i\left(\overline{a}_+ - \overline{a}^\dagger_+\right)
- 2\epsilon\delta\, F_\epsilon(N)\, \overline{\sigma}_1\Big],
\end{equation}
while the projected conjugate momenta are such that,
\begin{equation}
\overline{p}_i+\frac{1}{2}B\epsilon_{ij}\overline{x}_j=-\epsilon\delta\sqrt{2\hbar B}\, F_\epsilon(N)\, \overline{\sigma}_i.
\end{equation}
Hence, rather than vanish as in the ${\cal N}=0$ case, these specific combinations of projected bosonic
operators reduce purely to the bosonic-fermionic variables $\overline{\sigma}_i$. Thus from this point of
view one may consider the projected Cartesian coordinates, $\overline{x}_i$, together with the bosonic-fermionic
spin variables, $\overline{\sigma}_i$, to provide the complete set of independent variables spanning the ${\cal N}=1$
supersymmetric Moyal--Voros superplane. Consequently, the only independent commutation relations still to
be considered are those among these quantities. One finds,
\begin{equation}
\Big[\overline{x}_i,\overline{x}_j\Big]=-\frac{i\hbar}{B}\epsilon_{ij}
\Big(\mathbb{P}_+\,-\,F^2_\epsilon(N)\,\mathbb{P}_-\Big),
\end{equation}
which thus specifies the noncommutative geometry of the superplane in its bosonic sector, while
\begin{equation}
\Big[\overline{x}_i,\overline{\sigma}_j\Big]=-i\epsilon\delta\sqrt{\frac{\hbar}{2B}} F_\epsilon(N)\delta_{ij}\mathbb{P}_-,
\end{equation}
specifies that noncommutativity between the bosonic and fermionic sectors, while the remaining
structures of (anti)commutators in (\ref{eq:fermion}) in the fermionic sector characterise the
mixed bosonic-fermionic character of the latter. Once again note the manifest SO(2) covariance properties of all these relations
for rotations in the plane.

\subsection{Covariance properties of ${\cal N}=1$ supersymmetric Moyal--Voros superplanes}

Besides the algebraic relations characterising the ${\cal N}=1$ supersymmetric Moyal--Voros superplane it is also of interest, because of its built-in supersymmetric covariance properties, to consider the
action of the projected ${\cal N}=1$ supercharge on the defining variables of the superplane.
In order to present results of later interest and of somewhat more general use, let consider a slight
generalisation of the operator $\overline{Q}_0$ in the form
\begin{equation}
Q_1=\sqrt{\hbar\omega_c}\Big(\alpha_+\mathbb{P}(N,\delta) + \alpha_-\mathbb{P}(N+1,\epsilon\delta)\Big),
\end{equation}
such that
\begin{equation}
Q^2_1=\hbar\omega_c\Big(\alpha^2_+\mathbb{P}(N,\delta) + \alpha^2_-\mathbb{P}(N+1,\epsilon\delta)\Big).
\end{equation}
With the choice $\alpha_+=\delta\sqrt{N}$ and $\alpha_-=\epsilon\delta\sqrt{N+1}$, $\overline{Q}_1$ corresponds to
the operator $Q_0$.

Since the right-handed chiral bosonic sector is ${\cal N}=1$ supersymmetric invariant, it follows that
\begin{equation}
\Big[Q_1,\overline{a}_+\big]=0,\qquad
\Big[Q_1,\overline{a}^\dagger_+\Big]=0.
\end{equation}
Once again both the commutation and anticommutation relations of the variables $\overline{\sigma}_\pm$ with the supercharge are
specified by the structure of the projection and their products with $Q_1$ are easily identified
\begin{equation}
Q_1\overline{\sigma}_\pm=\sqrt{\hbar\omega_c}\,\alpha_\pm\,\overline{\sigma}_\pm,\qquad
\overline{\sigma}_\pm Q_1=\sqrt{\hbar\omega_c}\,\alpha_\mp\,\overline{\sigma}_\pm.
\end{equation}
Similar relations hold for the products $Q_1\overline{\sigma}_i$ and $\overline{\sigma}_i Q_1$, which can easily be worked out from these. Consequently,
\begin{equation}
\Big[Q_1,\overline{\sigma}_\pm\Big]=\pm\sqrt{\hbar\omega_c}\,\left(\alpha_+ - \alpha_-\right)\,\overline{\sigma}_\pm,\qquad
\Big\{Q_1,\overline{\sigma}_\pm\Big\}=\sqrt{\hbar\omega_c}\,\left(\alpha_+ + \alpha_-\right)\,\overline{\sigma}_\pm,
\end{equation}
as well as
\begin{equation}
\Big[Q_1,\overline{\sigma}_i\Big]=i\sqrt{\hbar\omega_c}\,\left(\alpha_+ - \alpha_-\right)\,\epsilon_{ij}\,\overline{\sigma}_j,\qquad
\Big\{Q_1,\overline{\sigma}_i\Big\}=\sqrt{\hbar\omega_c}\,\left(\alpha_+ + \alpha_-\right)\,\overline{\sigma}_i.
\end{equation}
It then also follows that for the Cartesian superplane coordinates,
\begin{equation}
\Big[Q_1,\overline{x}_i\Big]=-\frac{i\hbar}{\sqrt{2m}}\,2\epsilon\delta\,F_\epsilon(N)\,\left(\alpha_+ - \alpha_-\right)\,
\overline{\sigma}_i,
\end{equation}
showing that indeed under the ${\cal N}=1$ supersymmetry transformations these variables are mapped into the
spin degrees of freedom, while the latter, due to their dual bosonic-fermionic character are simply mapped back
into themselves, in a manner still consistent with supersymmetry.

Under SO(2) planar rotations, the projected variables of the Moyal--Voros superplane transform according to the relations,
\begin{equation}
\Big[\overline{L},\overline{a}^\dagger_+\Big]=\hbar\,\overline{a}^\dagger_+,\qquad
\Big[\overline{L},\overline{a}_+\Big]=-\hbar\,\overline{a}_+,\qquad
\Big[\overline{L},\overline{\sigma}_\pm\Big]=\pm\hbar\,\overline{\sigma}_\pm,\qquad
\Big[\overline{L},\overline{\sigma}_i\Big]=i\hbar\,\epsilon_{ij}\,\overline{\sigma}_j.
\end{equation}
These properties are identical to the corresponding commutation relations before projection (because $L$ commutes
with the considered projection). Hence, indeed, the applied projection remains manifestly covariant under both ${\cal N}=1$
supersymmetry and SO(2) rotations in the Euclidean plane.

\section{Conclusions}

By considering the ${\cal N}=1$ supersymmetric Landau problem and non trivial projections onto its Landau levels
consistent with supersymmetry, a ${\cal N}=1$ supersymmetric covariant analogue of the ordinary
noncommutative Moyal--Voros plane has been identified, independently of the choice of signs $(\delta,\epsilon)$
and involving some normalisation factors dependent on the choice of Landau levels $N$ and $(N+1)$ onto which the projection
is performed. To perhaps better make manifest the general structure that has thereby emerged, it is useful
to apply the following changes of notation,
\begin{eqnarray}
|n_+,N;\delta\rangle \rightarrow |n;\tau=+1\rangle &,&
|n_+,N+1;\epsilon\delta\rangle \rightarrow |n;\tau=-1\rangle, \nonumber \\
\mathbb{P}(N,\delta) \rightarrow \mathbb{P}(\tau=+1) &,&
\mathbb{P}(N+1,\epsilon\delta) \rightarrow \mathbb{P}(\tau=-1), \nonumber \\
\mathbb{P}_+ \rightarrow \mathbb{E}=\mathbb{P}(+1)+\mathbb{P}(-1) &,&
\mathbb{P}_- \rightarrow \tau_3=\mathbb{P}(+1)-\mathbb{P}(-1).
\end{eqnarray}
In this notation
\begin{eqnarray}
\overline{a}_+ \rightarrow b=\sum_{\tau=\pm 1}\sum_{n=0}^\infty |n;\tau\rangle\,\sqrt{n+1}\,\langle n+1;\tau| &,&
\overline{a}^\dagger_+ \rightarrow b^\dagger=\sum_{\tau\pm 1}\sum_{n=0}^\infty |n+1;\tau\rangle\,\sqrt{n+1}\,\langle n;\tau| , \nonumber \\
-2i\epsilon\delta\,\overline{\sigma}_+ \rightarrow \tau_+=\sum_{n=0}^\infty |n;+1\rangle\langle n;-1| &,&
+2i\epsilon\delta\,\overline{\sigma}_- \rightarrow \tau_-=\sum_{n=0}^\infty |n;-1\rangle\langle n;+1| .
\end{eqnarray}
It is also useful to define the following combinations,
\begin{equation}
\tau_1=i\left(\tau_+ - \tau_-\right) = 2\epsilon\delta\,\overline{\sigma}_1,\qquad
\tau_2=\tau_+ + \tau_- = 2\epsilon\delta\,\overline{\sigma}_2.
\end{equation}
Instead of the $(b,b^\dagger)$ Fock operators, let us rather introduce a Cartesian ${\cal N}=0$
Moyal--Voros plane parametrisation in the form,
\begin{equation}
u_1=\sqrt{\frac{\hbar}{2B}}\left(b+b^\dagger\right),\qquad
u_2=\sqrt{\frac{\hbar}{2B}}\,i\left(b-b^\dagger\right).
\end{equation}
Finally, let us introduce the real dimensionless parameter $\lambda$ in place of the quantity $F_\epsilon(N)$.

In terms of these new notations, the structure of the ${\cal N}=1$ non(anti)commutative Moyal--Voros superplane
is as follows. It is spanned by the operators $\overline{x}_i$ and $\tau_i$ ($i=1,2$), $\overline{x}_i$ being the
Cartesian bosonic variables and $\tau_i$ being bosonic-fermionic spin degrees of freedom, with the
parametrisation\footnote{Note that $\sqrt{\hbar/(2B)}$ is the magnetic length of the Landau problem.}
\begin{equation}
\overline{x}_i=u_i+\lambda\sqrt{\frac{\hbar}{2B}}\,\epsilon_{ij}\,\tau_j.
\label{eq:solder}
\end{equation}
The algebra of these operators is characterised by the following (anti)commutation relations,
\begin{equation}
\Big[u_i,u_j\Big]=-\frac{i\hbar}{B}\,\mathbb{E},
\end{equation}
\begin{equation}
\tau^2_+=0,\qquad\tau^2_-=0,\qquad
\Big\{\tau_+,\tau_-\Big\}=\mathbb{E},\qquad
\Big\{\tau_i,\tau_j\Big\}=2\delta_{ij}\,\mathbb{E},
\end{equation}
\begin{equation}
\Big[\tau_+,\tau_-\Big]=\tau_3,\qquad
\Big[\tau_i,\tau_j\Big]=2i\epsilon_{ij}\,\tau_3.
\end{equation}
Consequently,
\begin{equation}
\Big[\overline{x}_i,\overline{x}_j\Big]=-\frac{i\hbar}{B}\,\epsilon_{ij}\,\left(\mathbb{E}\,-\,\lambda^2\,\tau_3\right),\qquad
\Big[\overline{x}_i,\tau_j\Big]=-2i\lambda\sqrt{\frac{\hbar}{2B}}\,\delta_{ij}\,\tau_3.
\end{equation}
Note that the bosonic-fermionic spin degrees of freedom sector is such that the only possible representation of both the anticommutation relations and SU(2) commutation relations of the Clifford algebra is in terms of the
$2\times 2$ Pauli matrices, in the case of the present ${\cal N}=1$ supersymmetry construction.
What is perhaps even more intriguing is that the bosonic sector of the superplane coordinates, $\overline{x}_i$,
is realised through a soldering with the spin degrees of freedom $\tau_i$ of two copies---distinguished by
the eigenvalues $\tau=\pm 1$ of $\tau_3$---of the ordinary ${\cal N}=0$ noncommutative Moyal--Voros plane spanned by $u_i$,
while the free parameter $\lambda$ sets the strength of that soldering. It is as if the ordinary Moyal--Voros plane
had been ``fattened" by soldering onto it a spin 1/2 structure, while at the same time making more fuzzy the notion
of a bosonic or a fermionic variable since $\tau_i$ possess both characters while $\overline{x}_i$ is constructed as a linear
combination out of these as well as a bosonic Fock algebra.

Considering a supersymmetry-like charge,
\begin{equation}
Q_1=\sqrt{\hbar\omega_c}\Big[\alpha_+ \mathbb{P}(+1) + \alpha_- \mathbb{P}(-)\Big],
\end{equation}
$\alpha_\pm$ being two real arbitrary constants, which leaves all states $|n;\tau\rangle$ invariant, and for which we thus have
\begin{equation}
Q^2_1=\hbar\omega_c\Big[\alpha^2_+ \mathbb{P}(+1) + \alpha^2_- \mathbb{P}(-)\Big],
\end{equation}
its action on the superplane supercoordinates is such that,
\begin{equation}
\Big[Q_1,u_i\Big]=0,\qquad
\Big[Q_1,\overline{x}_i\Big]=-i\frac{\hbar}{\sqrt{2m}}\,\lambda\,\left(\alpha_+ - \alpha_-\right)\,\tau_i,
\end{equation}
and
\begin{equation}
\Big[Q_1,\tau_\pm\Big]=\pm\sqrt{\hbar\omega_c}\,\left(\alpha_+ - \alpha_-\right)\,\tau_\pm,\qquad
\Big\{Q_1,\tau_\pm\Big\}=\sqrt{\hbar\omega_c}\,\left(\alpha_+ + \alpha_-\right)\,\tau_\pm,
\end{equation}
\begin{equation}
\Big[Q_1,\tau_i\Big]=i\sqrt{\hbar\omega_c}\,\left(\alpha_+ - \alpha_-\right)\,\epsilon_{ij}\tau_j,\qquad
\Big\{Q_1,\tau_i\Big\}=\sqrt{\hbar\omega_c}\,\left(\alpha_+ + \alpha_-\right)\,\tau_i.
\end{equation}
In the Landau problem context one has $\alpha_+=\delta\sqrt{N}$ and $\alpha_-=\epsilon\delta\sqrt{N+1}$
with $\lambda=F_\epsilon(N)$ ($N\ge 1$). However, in a more general context, given only the two sector structure
of the representation space of the ${\cal N}=1$ Moyal--Voros superplane in the quantum number $\tau=\pm 1$, one is free to choose
these three real parameters, with in particular $Q^2_1$ then playing the r\^ole of a Hamiltonian operator for the modelling of some
physical system. For instance by choosing $\alpha_+=\alpha_-=\alpha$, the two sectors, $\tau=\pm 1$, are degenerate
in energy and SO(2) covariant with states which are all eigenstates of $Q_1$ with a common eigenvalue
$\alpha\sqrt{\hbar\omega_c}$, while all supercoordinates $\overline{x}_i$ and $\tau_i$ have vanishing
commutators with $Q_1$.

However, it remains to be seen whether such a ${\cal N}=1$ Moyal--Voros superplane framework is of
any relevance to the quantum Hall problem, be it in its integer or fractional realisations.
Furthermore, the above picture of two ordinary Moyal--Voros planes soldered by the bosonic-fermionic
spin degrees of freedom is strangely reminiscent of two $D$-branes stacked on top of one another\cite{Witten}
in a limit leading to noncommutativity in M-theory\cite{SW}. It may be worth understanding the possible relation
between these two situations, if any.

Besides such wider ranging physical issues, given the present construction, one interesting task remaining
to be completed now is the identification of a Grassmann graded $\star$-product defined on functions
of a ${\cal N}=1$ supersymmetric extension of the commutative Cartesian coordinates, $(x_1,x_2)$, of the plane\cite{scholtz2}. Beyond that, the generalisation of the present approach by including a larger number of supersymmetries
is certainly worthwhile considering in order to enrich the collection of such Moyal--Voros superplanes beyond
the ${\cal N}=1$ case\cite{BenGeloun2}.

\section*{Acknowledgements}

Part of this work was completed during the International Workshop on Coherent States, Path Integrals and
Noncommutative Geometry, held at the National Institute for Theoretical Physics (NITheP, Stellenbosch, South
Africa) on 4--22 May 2009. J.G. wishes to thank the organisers as well as NITheP for the financial support
having made his participation possible, and for NITheP's always warm, inspiring and wonderful hospitality.

The work of J.B.G. and F.G.S. is supported under a grant of the National Research Foundation of South Africa.
J.G. acknowledges the Abdus Salam International Centre for Theoretical Physics (ICTP, Trieste, Italy)
Visiting Scholar Programme in support of a Visiting Professorship at the ICMPA-UNESCO (Republic of Benin).
The work of J.G. is supported in part by the Institut Interuniversitaire des Sciences Nucl\'eaires (I.I.S.N., Belgium),
and by the Belgian Federal Office for Scientific, Technical and Cultural Affairs through
the Interuniversity Attraction Poles (IAP) P6/11.

\end{document}